\documentclass[aps,pra,preprintnumbers,showpacs,tightenlines]{revtex4}

\usepackage{amssymb}
\usepackage{amsmath}
\usepackage{graphicx}
\usepackage{epsfig}
\usepackage{subfigure}
\usepackage{amsfonts}
\usepackage{CJK}

\begin{document}

\title{Proposal for implementing the three-qubit refined Deutsch-Jozsa quantum algorithm}

\author{Qi-Ping Su$^{1}$, Man Liu$^{1}$, and Chui-Ping Yang$^{1,2}$\thanks{E-mail: yangcp@hznu.edu.cn}}
\thanks{E-mail: yangcp@hznu.edu.cn}
\address{$^1$Department of Physics, Hangzhou Normal University,
Hangzhou, Zhejiang 310036, China}

\address{$^2$State Key Laboratory of Precision Spectroscopy, Department of Physics,
East China Normal University, Shanghai 200062, China}

\date{\today}

\begin{abstract}
We propose a way to implement a three-qubit refined Deutsh-Jozsa (DJ)
algorithm. The present proposal is based on the construction
of the 35 $f$-controlled phase gates, which uses single-qubit $\sigma_z$
gates and two-qubit {\it standard} controlled-phase (CP) gates only. This proposal
is implementable because a single-qubit $\sigma_z$ gate can be easily realized
by applying a single classical pulse and a two-qubit CP gate has been
experimentally demonstrated in various physical systems. Finally, it is
noted that this proposal is quite general, and can be applied to implement
a three-qubit refined DJ algorithm in a cavity-based or noncavity-based
physical system.
\end{abstract}

\maketitle
\date{\today}

\section{Introduction}

The interest in quantum computation is stimulated by the discovery of
quantum algorithms which can solve problems of significance much more
efficiently than their classical counterparts. Quantum algorithms could be
implemented using various types of qubits [1] or quantum simulators [2].
Among important quantum algorithms, there exist the Deutsch algorithm [3],
the Deutsch-Jozsa algorithm [4], the Shor algorithm [5], the Simon algorithm
[6], the quantum Fourier transform algorithm, and the Grover search
algorithm [7]. In addition, proposals for implementing other quantum algorithms
have been also presented (e.g., see [8]).

As is well known, the Deutsch algorithm and the Deutsch-Jozsa
algorithm were the first two that make use of the features of quantum
mechanics for quantum computation. Compared with other quantum algorithms,
these two algorithms are easy to be implemented and thus have been
considered as the natural candidates for demonstrating power of quantum
computation.

In this article, we restrict ourself to a refined Deutsch-Jozsa (DJ)
algorithm. We will propose a way to implement a three-qubit \textit{refined}
DJ algorithm. As shown below, this proposal is based on the construction of
the 35 $f$-controlled phase gates, which employs single-qubit $\sigma _z$
gates and two-qubit \textit{standard} controlled-phase (CP) gates. To the
best of our knowledge, our work is the first to consider how to construct
the 35 $f$-controlled phase gates using these two kinds of basic gates. The
construction of the 35 $f$-controlled phase gates here is general, and can
be applied to implement the three-qubit refined DJ algorithm in a
cavity-based or noncavity-based physical system.

\section{Motivations}

There are several motivations for this work:

(i) In Ref.~[9], the authors showed that implementing a one-qubit or
two-qubit \textit{refined} DJ algorithm does not require entanglement
between the input query qubits, and thus test of the one-qubit or two-qubit
\textit{refined} DJ algorithm is not meaningful because it can be solved in
a classical way. However, they showed that entanglement exists between the
input query qubits during performing a $n$-qubit refined DJ algorithm with $%
n\geq 3$, and thus the meaningful test of the refined DJ algorithm occurs
only and only if $n\geq 3$. Hence, as far as the refined DJ algorithm is
concerned, a three-qubit refined DJ algorithm (i.e., the case for $n=3$) is
the smallest one that needs to be tested or implemented, in order to
demonstrate the full power of quantum computation as applied to the Deutsch
problem. For a detailed discussion, see [9].

(ii) A second motivation of this work is as follows. A single-qubit $\sigma
_{z}$ gate can be easily realized by applying a single classical pulse. In
addition, it is well known that a two-qubit CP gate (i.e., the key element
in the construction of the 35 $f$-controlled phase gates, see Table II
below) has been experimentally demonstrated in many physical systems such as
quantum dots, trapped ions, atoms in cavity QED, and superconducting
qubits/qutrits coupled to a single circuit cavity. Thus, it is
straightforward to implement a three-qubit refined DJ algorithm by applying
the present proposal.

(iii) Quantum information processing using superconducting qubits coupled to
a cavity has attracted considerable interest during the past ten years.
Based on cavity QED, many theoretical proposals have been presented for
realizing two-qubit gates [10-14] and multiple qubit gates [15,16] with
superconducting qubits. Moreover, experimental demonstration of two-qubit
gates [17-20] and three-qubit gates [21-23] with superconducting qubits in
cavity QED has been also reported. However, after a deep investigation, we
noted that how to implement a three-qubit (original and refined) DJ
algorithm with superconducting qubits or qutrits in cavity QED has not been
reported in both theoretical and experimental aspects.~As is known, the
experimental realization of the three-qubit DJ algorithm with a
cavity-superconducting-device system is important because it would be an
important step toward more complex quantum computation in circuit cavity QED.

(iv) Over the past decade, there has been much interest in quantum
information processing with atoms in cavity QED. Based on cavity QED
technique, many theoretical methods have been proposed for implementing a
two-qubit CP or controlled-NOT gate and multiple qubit controlled gates with
atoms [24-28]. Moreover, a two-qubit CP gate between a cavity mode and an
atom has been experimentally demonstrated [29]. However, after a thorough
search, we found that only several proposals [30-32] were proposed for
implementing the original or refined $n$-qubit DJ algorithm for $n\geq 3$,
using atoms in cavity QED. As discussed below, these proposals have some
drawbacks. For instances, the one in [30] is incomplete, and the ones in
[31,32] are difficult to implement in experiments when compared with our
current proposal.

\section{Deutsch-Jozsa Algorithm}

The DJ algorithm is aimed at distinguishing the constant function from the
balanced functions on $2^{n}$ inputs. The function $f\left( x\right) $ takes
either $0$ or $1.$ For the constant function, the function values are
constant ($0$ or $1$) for all $2^{n}$ inputs. In contrast, for the balanced
function, the function values are equal to 1 for half of all the possible
inputs, and $0$ for the other half. Using the DJ algorithm, whether the
function is constant or balanced can be determined by only one query.
However, a classical algorithm would require $2^{n-1}+1$ queries to answer
the same problem, which grows exponentially with input size.

The \textit{original} DJ algorithm involves the $n$ input query qubits each
initially prepared in the state $\left\vert 0\right\rangle $ and an
auxiliary working qubit initially prepared in the state\ $\left\vert
-\right\rangle =\left( \left\vert 0\right\rangle -\left\vert 1\right\rangle
\right) /\sqrt{2}$. It operates by: (i) applying a Hadamard gate on each
input query qubit, which results in $\left\vert 0\right\rangle \rightarrow
\left( \left\vert 0\right\rangle +\left\vert 1\right\rangle \right) /\sqrt{2}
$ and $\left\vert 1\right\rangle \rightarrow \left( \left\vert
0\right\rangle -\left\vert 1\right\rangle \right) /\sqrt{2}$), (ii)
performing a $f$-controlled-NOT gate $\widetilde{U}_{f},$ and (iii) applying
a Hadamard gate again on each input query qubit followed by a measurement on
each input query qubit along the single-qubit $z$ base formed by the two
states $\{\left\vert 0\right\rangle ,\left\vert 1\right\rangle \}.$ Here, $%
\widetilde{U}_{f}$ is defined as $\widetilde{U}_{f}\left\vert x\right\rangle
\left\vert y\right\rangle =\left\vert x\right\rangle \left\vert y\oplus
f\left( x\right) \right\rangle ,$ where $\left\vert x\right\rangle $ is the
state of the $n$ input query qubits with $x$ being an $n$-qubit binary
decomposition, $\left\vert y\right\rangle $ is the state of the working
qubit with $y\in \{0,1\},$ and $\oplus $ means modulo 2. Note that the
operation $\widetilde{U}_{f}$ results in the following transformation [2]
\begin{equation}
\left\vert x\right\rangle \left\vert -\right\rangle \rightarrow \left(
-1\right) ^{f\left( x\right) }\left\vert x\right\rangle \left\vert
-\right\rangle ,
\end{equation}%
which shows that there is no entanglement between the query qubits and the
working qubit in the output of the $\widetilde{U}_{f}$. Thus there is no
need for any coupling between the query qubits and the working qubit during
the entire process. Hence, in the \textit{original} DJ algorithm the working
qubit is completely redundant.

\begin{figure}[tbp]
\begin{center}
\includegraphics[bb=134 565 467 637, width=10.5 cm, clip]{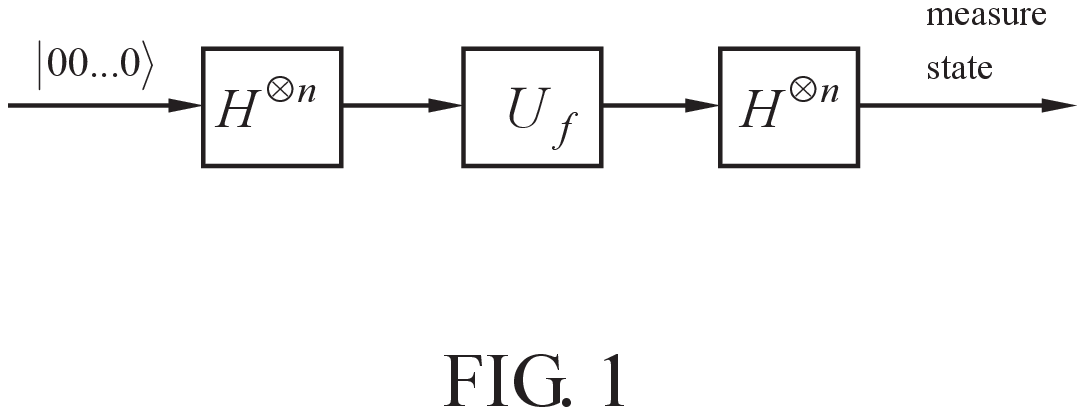} \vspace*{%
-0.08in}
\end{center}
\caption{Quantum circuit for the refined $n$-qubit Deutsch-Jozsa algorithm. Here, $H$ represents
a single-qubit Hadamard gate on an input query qubit, which results in $\left\vert 0\right\rangle \rightarrow
\left( \left\vert 0\right\rangle +\left\vert 1\right\rangle \right) /\sqrt{2}
$ and $\left\vert 1\right\rangle \rightarrow \left( \left\vert
0\right\rangle -\left\vert 1\right\rangle \right) /\sqrt{2}$). In addition, $U_{f}$ indicates
a $f$-controlled phase gate on the $n$ input query qubits, described by Eq.~(2).}
\label{fig:1}
\end{figure}

The \textit{refined} DJ algorithm was proposed by Collins \textit{et al.} in
2001 [9], which is illustrated in Fig.~1. This refined DJ algorithm follows
a similar pattern of operations as the original DJ algorithm, which is
described below:

(i) Each input query qubit is prepared in the initial state $\left|
0\right\rangle .$

(ii) Perform a Hadamard transformation $H$ on each qubit. As a result, the $n
$-qubit initial state $\left\vert 00\cdot \cdot \cdot 0\right\rangle $
changes to the state $\frac{1}{2^{n/2}}\sum_{x=0}^{2^{n}-1}\left\vert
x\right\rangle $ (denoted as $\left\vert \psi _{1}\right\rangle $)$.$

\begin{table}[tbp]
\begin{center}
\includegraphics[bb=0 0 800 800, width=10.0 cm, clip]{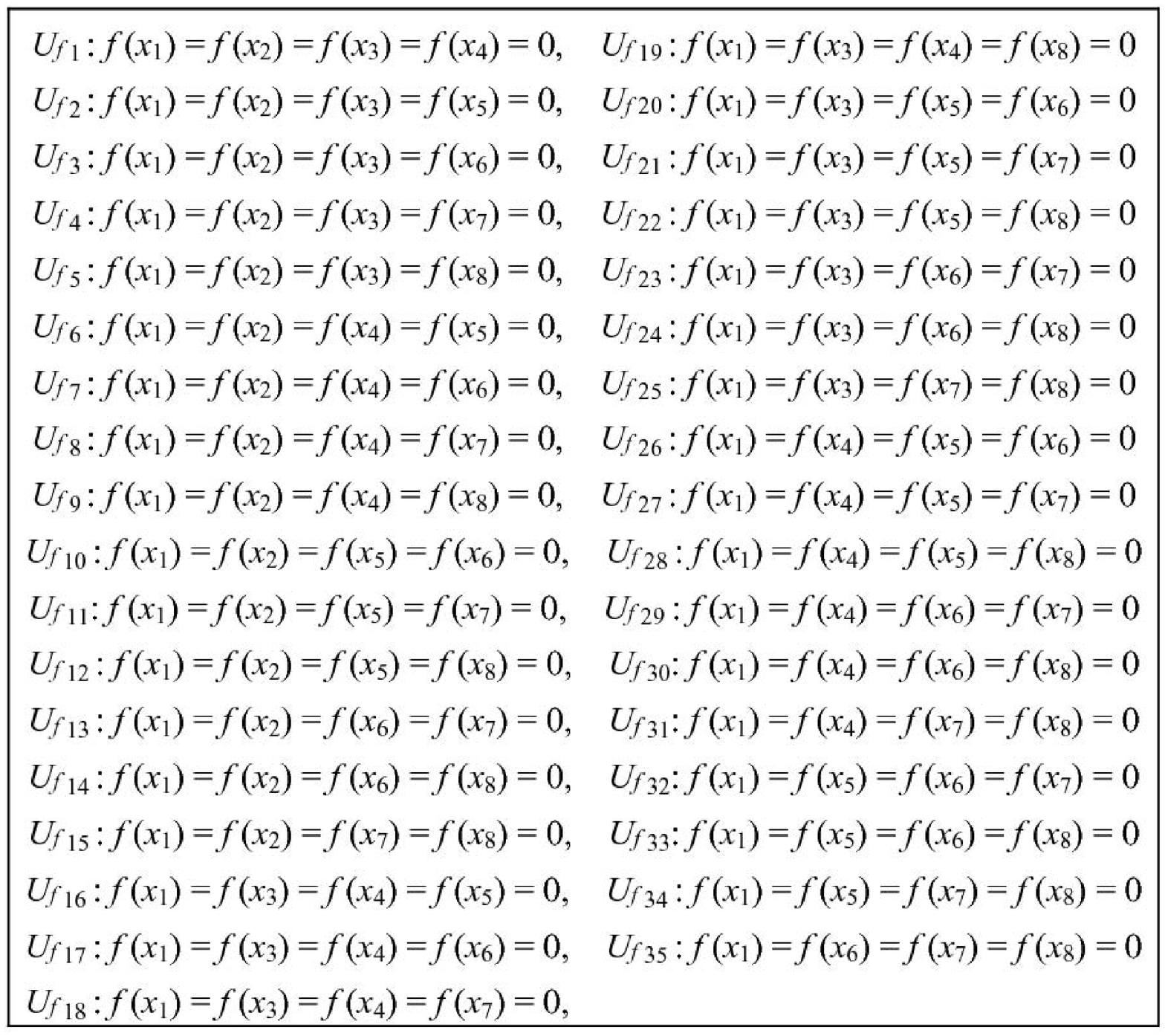}
\vspace*{-0.08in}
\end{center}
\caption{List of the 35 balanced functions for a three-qubit refined
Deutsh-Jozsa algorithm. Here, $%
x_{1}=000,x_{2}=001,x_{3}=010,x_{4}=011,x_{5}=100,x_{6}=101,x_{7}=110, $ and
$x_{8}=111.$ For simplicity, we only list the funtion values, which are $%
``0" $, for four inputs corresponding to each balanced function. Note that
for each balanced function, the function values for the other four inputs
(not listed) take a value $``1"$. For instance, for the balanced function
corresponding to $U_{f1},$ the function values for the other four inputs
(not listed) are $f\left( x_{5}\right) =f\left( x_{6}\right) =f\left(
x_{7}\right) =f\left( x_{8}\right) =1.$}
\label{table:1}
\end{table}

\begin{table}[tbp]
\begin{center}
\includegraphics[bb=0 0 900 900, width=10.0 cm, clip]{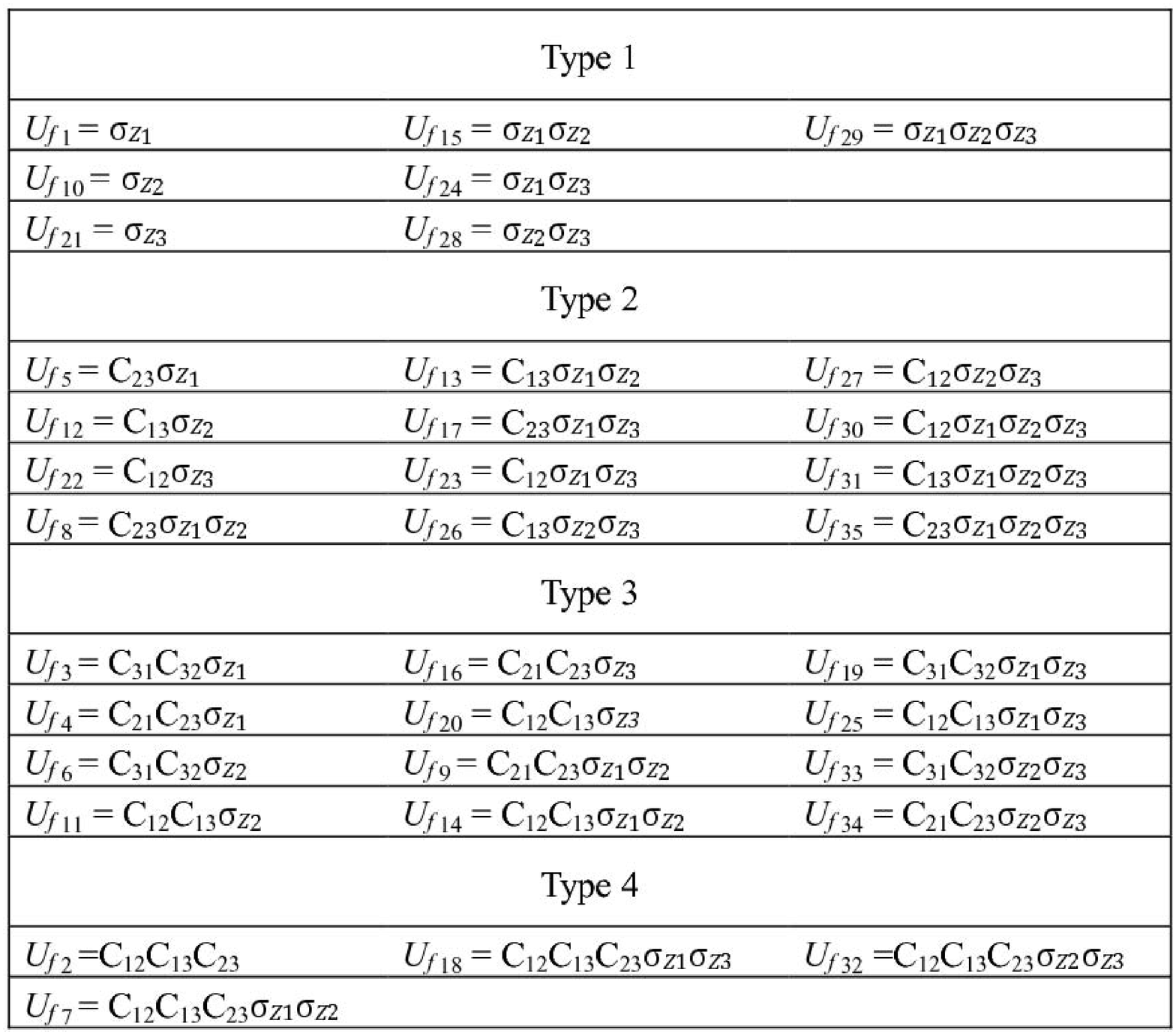}
\vspace*{-0.08in}
\end{center}
\caption{The construction of 35 $f$-controlled phase gates for a three-qubit refined
Deutsh-Jozsa algorithm. Here, $\protect\sigma _{z_j}$ represents a
single-qubit $\protect\sigma _{z}$ gate on qubit $j$ ($j=1,2,3$); while $%
C_{jk}$ indicates a two-qubit \textit{standard} controlled-phase gate on
qubits $j$ and $k$ ($j,k=1,2,3$), described by Eq. (3).}
\label{table:2}
\end{table}

(iii) Apply the $f$-controlled phase shift $U_{f}$, described by
\begin{equation}
\left\vert x\right\rangle \overset{U_{f}}{\longrightarrow }\left( -1\right)
^{f\left( x\right) }\left\vert x\right\rangle ,
\end{equation}
which leads the state $\left\vert \psi _{1}\right\rangle $ to the state $%
\frac{1}{2^{n/2}}\sum_{x=0}^{2^{n}-1}\left( -1\right) ^{f\left( x\right)
}\left\vert x\right\rangle $ (denoted as $\left\vert \psi _{2}\right\rangle $%
).

(iv) Perform another Hadamard transformation $H$ on each qubit. As a result,
the state $\left\vert \psi _{2}\right\rangle $ becomes $\frac{1}{2^{n}}%
\sum_{z=0}^{2^{n}-1}\sum_{x=0}^{2^{n}-1}\left( -1\right) ^{x\cdot z+f\left(
x\right) }\left\vert z\right\rangle .$

(v) Measure the final state of the $n$ qubits. If the $n$ qubits are
measured in the state $\left\vert 00...0\right\rangle ,$ then $f\left(
x\right) $ is constant. However, if they are measured in other $n$-qubit
computational basis states, then $f\left( x\right) $ is balanced. This is
because the amplitude $a_{\left\vert 00...0\right\rangle }$ of the state $%
\left\vert 00...0\right\rangle $ is given by $a_{\left\vert
00...0\right\rangle }=\frac{1}{2^{n}}\sum_{x=0}^{2^{n}-1}\left( -1\right)
^{f\left( x\right) },$ which is $\pm 1$ for a constant $f\left( x\right) $
while $0$ for a balanced $f\left( x\right) $.

One can see that when compared with the original DJ algorithm, this refined
DJ algorithm does not need a working qubit. Hence, it requires one qubit
fewer than the original DJ algorithm. Consequently, its physical
implementation requires one fewer two-state system.

\section{Construction of the $f$-controlled phase gates}

For both of the original and refined DJ algorithms, there are a total of $%
C_{2^{n}}^{2^{n-1}} $ balanced functions, among which only $%
C_{2^{n}}^{2^{n-1}}/2$ balanced functions are nontrivial if the symmetry is
taken into account. For the three-qubit DJ algorithm, i.e., $n=3,$ there
thus exist $C_{8}^{4}/2=35$ nontrivial balanced functions $%
U_{f1},U_{f2},...,U_{f35}$ (see Table I).

A single qubit $\sigma _z$ gate results in the transformation $\sigma
_z\left| 0\right\rangle =\left| 0\right\rangle $ while $\sigma _z\left|
1\right\rangle =-\left| 1\right\rangle .$ As is well known, a two-qubit
\textit{standard} CP gate $C_{jk}$ on qubits $j$ and $k$ is described as
follows
\begin{eqnarray}
\left| 00\right\rangle _{jk} &\rightarrow &\left| 00\right\rangle
_{jk},\;\;\left| 10\right\rangle _{jk}\rightarrow \left| 10\right\rangle
_{jk},  \notag \\
\left| 01\right\rangle _{jk} &\rightarrow &\left| 01\right\rangle
_{jk},\;\;\left| 11\right\rangle _{jk}\rightarrow -\left| 11\right\rangle
_{jk},
\end{eqnarray}
which implies that if and only if the control qubit $j$ (the first qubit) is
in the state $\left| 1\right\rangle $ , a phase flip happens to the state $%
\left| 1\right\rangle $ of the target qubit $k$ (the second qubit), but
nothing happens otherwise.

The construction for each of the 35 $f$-controlled phase gates is listed in
Table II. One can see from Table II that the 35 $f$-controlled phase gates $%
U_{f1},U_{f2},...,U_{f35}$ are classified into the following four types: (i)
Type 1 includes seven $f$-controlled phase gates each constructed with
single-qubit $\sigma _{z}$ gates only; (ii) Type 2 contains twelve\ $f$%
-controlled phase gates each constructed with single-qubit $\sigma _{z}$
gates and one two-qubit CP gate; (iii) Type 3 has twelve $f$-controlled
phase gates each constructed by using single-qubit $\sigma _{z}$ gates and
two two-qubit CP gates; and (iv)\ Type 4 involves four $f$-controlled phase
gates each implemented with two single-qubit $\sigma _{z}$ gates and three
two-qubit CP gates.

As discussed previously, a single-qubit $\sigma _{z}$ gate here can be
easily realized by applying a single classical pulse, and a two-qubit CP
gate (i.e., the central element in our gate construction) has been
experimentally demonstrated in various physical systems. Hence, the gate
construction given in Table II is implementable.

\section{Comparing with previous proposals using atoms in cavity QED}

After a thorough search, we found that only several proposals [30-32] were
proposed for implementing the original or refined DJ algorithm for $n\geq 3$%
, using atoms in cavity QED. In the following we will briefly introduce
these previous proposals and then give a comparison of them with our current
proposal.

(i) The authors in [30] proposed an \textit{incomplete}\ scheme for
realizing the $n$-qubit original DJ algorithm using atoms coupled to a
cavity, because they only discussed how to implement $one$ balanced
function, which completes the state transformation from the input state $%
2^{-n/2}\left( \left\vert 0\right\rangle +\left\vert 1\right\rangle \right)
^{\otimes n}$ to the output state $2^{-n/2}\left( \left\vert 0\right\rangle
-\left\vert 1\right\rangle \right) ^{\otimes n}$, as shown in [30].
Furthermore, this state transformation can be reached in a classical way,
since it can be alternatively achieved by simply performing a single-qubit $%
\sigma _{z}$ gate on each of individual qubits of the register.

(ii) In Ref. [31], a scheme was proposed for the implementation of the $n$%
-qubit original DJ algorithm via atoms interacting with a cavity. As
discussed there, to implement each of $C_{2^{n}}^{2^{n-1}}/2$ balanced
functions, this scheme requires $2^{n-1}$ different $n$-qubit controlled-NOT
gates each with $n-1$ qubits simultaneously controlling a target qubit. For
a three-qubit DJ algorithm, $n=3.$ Thus, by applying this scheme to realize
each balanced function for a three-qubit DJ algorithm, four different
three-qubit controlled-NOT gates (each with 2 qubits simultaneously
controlling a target qubit) would be required.

(iii) Ref. [32] presented a way for implementing a 3-qubit refined DJ
algorithm with three atoms in cavity QED. This proposal is based on four
three-qubit controlled phase gates (each with 2 qubits simultaneously
controlling a target qubit). In addition, as discussed in [32], it is
required to send three atoms through four cavities at the same time, or
simultaneously sending the three atoms through a common cavity four times.

From the description given above, one can see that:

(i) The present proposal is different from the previous one in [30], since
the latter is \textit{incomplete}, which only considered realizing one
balanced function that can be achieved in a classical way. In contrast to
[30], we have presented a \textit{complete} protocol for implementing all 35
$f$-controlled phase gates for a three-qubit refined DJ algorithm.

(ii) To implement a three-qubit DJ algorithm, the proposal in [31] requires
using four different three-qubit gates, and the one in [32] requires sending
three atoms through a cavity four times or needs the use of four cavities.
In contrast, it can be seen from Table II that our proposal needs three
two-qubit CP gates at most, which can be implemented using a single cavity
or resonator only, as discussed below. Hence, when compared with the
previous proposals in [31,32], it is simple and easy to implement a
three-qubit refined DJ algorithm based on cavity QED, by using the present
proposal.

We point out that it is not our intention to cast aspersions on existing
approaches [30-32] to the DJ algorithm implementation; rather we simply wish
to present a protocol to implement a three-qubit refined DJ algorithm.

As relevant to the cavity QED-based implementation of the present proposal,
some points may need to be addressed here. First, the present proposal can
be implemented with three superconducting qubits/qutrits placed in a single
cavity or resonator. This is because during performing each two-qubit CP
gate one can have the irrelevant qubit/qutrit to be decoupled from the
cavity by adjusting its level spacings (e.g., see, [15,16] and discussion
therein). Second, it can be realized using three atoms plus a single cavity.
During performing each two-qubit CP gate on atoms, the irrelevant atom can
be made to be decoupled from the cavity, for instance, by moving it out of
the cavity via translating optical lattices (e.g., see [16,28]).

\section{Conclusion}

We have proposed a way to implement the three-qubit refined Deutsch-Jozsa
quantum algorithm. The construction of the $f$-controlled phase gates can be
applied to implement a three-qubit refined Deutsch-Jozsa algorithm in a
cavity-based or noncavity-based physical system. This work is of interest
because it provides a simple and general protocol for implementing a
three-qubit refined Deutsch-Jozsa algorithm, which is an important step
toward more complex quantum computation.

\begin{center}
\textbf{ACKNOWLEDGMENTS}
\end{center}

C.P.Y. was supported in part by the National Natural Science Foundation of China under Grant No. 11074062, the Zhejiang
Natural Science Foundation under Grant No. LZ13A040002, the Open Fund from the
SKLPS of ECNU, and the funds from Hangzhou Normal University under Grant No.
HSQK0081. Q.P.S. was supported by Zhejiang Provincial Natural Science Foundation of China (Grant No. LQ12A05004).

\end{document}